\documentclass[11pt,twoside]{article}
\usepackage{jltp}
\usepackage{graphicx}

\title{Symmetries of Pairing Correlations in Superconductor-Ferromagnet Nanostructures}

\author{
M.~Eschrig,$^1$
\address{(1) Institut f{\"u}r Theoretische Festk{\"o}rperphysik and
DFG-Center for Functional Nanostructures,
Universit{\"a}t Karlsruhe, D-76128 Karlsruhe, Germany\\
(2) Forschungszentrum Karlsruhe, Institut f\"ur Nanotechnologie\\
D-76021 Karlsruhe, Germany}
T.~L\"ofwander,$^1$
T.~Champel,$^{1,*}$
J.~C.~Cuevas,$^{1,2,\dagger}$
J.~Kopu,$^{1}$
and~Gerd~Sch\"on$^{1,2}$
}

\runninghead{M. Eschrig {\it et al.}}{Symmetries of Pairing Correlations in
S-F Nanostructures}

\begin{document}

\begin{abstract}

\vspace{-2cm}

\begin{center}
\mbox{Dated: October 7, 2006}
\end{center}

Using selection rules imposed by the Pauli principle, we classify
pairing correlations according to their symmetry properties with
respect to spin, momentum, and energy.  We observe that inhomogeneity
always leads to mixing of even- and odd-energy pairing components.  We
investigate the superconducting pairing correlations present near
interfaces between superconductors and ferromagnets, with focus on
clean systems consisting of singlet superconductors and
either weak or half-metallic ferromagnets.  Spin-active
scattering in the interface region induces all of the possible
symmetry components.  In particular, the long-range equal-spin pairing
correlations have odd-frequency s-wave and even-frequency p-wave
components of comparable magnitudes. We also analyze the Josephson
current through a half-metal. We find analytic expressions and a
universality in the temperature dependence of the critical
current in the tunneling limit.  \\








\end{abstract}

\maketitle



\section{INTRODUCTION}

The rich physics specific for boundary regions between superconducting
and ferromagnetic materials has recently been probed in a series of
experiments
\cite{loehn05,beckmann04,per04,sch00,str94,kontos01,ryazanov01,blum02,
sellier04,bauer04,guichard03,lawrence96,giroud03,petrashov99,keizer06,
aumentado01,gu02,geers01,frolov04,Sta2006}.
Apart from confirming earlier theoretical predictions
\cite{bulaevskii77,buzdin82,bulaevskii85,rad1991} they provide deep
insight into the coexistence of the two types of order and have
inspired new ideas in the emerging field of spin electronics. In the
boundary region the characteristic correlations known from the
proximity effect in normal metals \cite{kogan82} are induced, but in
addition to the usual decay, they show an oscillating behavior
\cite{buzdin82}.  The length scales for decay and oscillations are set
by the magnetic length \cite{bulaevskii85}. It is smaller than the
decay length in a
normal metal, for ballistic systems by the ratio between temperature
and exchange energy, $k_BT/J$, and for diffusive systems by the factor
$\sqrt{k_BT/J}$.  This leads to modifications of the density of
states\cite{zar01,hal02,yok05,konstandin05} and the Josephson effect
through ferromagnets.\cite{gol04} For usual ferromagnets $J$ is
considerably larger than $k_BT$, and the proximity effect is hard to
observe.  A breakthrough came with the advance of dilute ferromagnetic
alloys with rather low spin polarization, such as Pd$_{1-x}$Ni$_x$ or
Cu$_{1-x}$Ni$_x$, where magnetic Ni ions are integrated into a
non-magnetic matrix. A high level of control has been reached with
heterostructures containing such ``weak'' ferromagnets 
(low spin polarization).
E.g., it became possible to spatially resolve properties
on the scale of the magnetic length and to observe the proximity
effect\cite{kontos01,ryazanov01}.

In parallel, it remained an important goal for the further development
of spin electronics to create and investigate
heterostructures\cite{lawrence96,giroud03,petrashov99,keizer06,aumentado01}
where ``strong'' ferromagnets with large exchange splitting of the
bands and high spin polarization are in contact with
superconductors. In the extreme case, for so-called half-metallic
materials, the polarization reaches values close to 100~\%. Half
metals are metallic in one spin direction with respect to a certain
spin quantization axis, and semiconducting or insulating in the other.
In the course of this research a so-called ``long-range proximity
effect'' was discovered
\cite{giroud03,petrashov99,keizer06,bergeret01,bergeret05,kadigrobov01,eschrig03,lofwander05},
which is governed by a length scale typical of the proximity effect in
a normal metal, rather than the magnetic length of the ferromagnet
involved. Such long-range proximity amplitudes have their origin in
the mixing between singlet and triplet pair amplitudes inherent to
pairing in the presence of broken spin-rotation
symmetry.\cite{bergeret05,eschrig03,champel05} Triplet correlations
are classified according to their projection on the spin quantization
axis that renders the quasiparticle bands in the ferromagnet diagonal.
There are three corresponding amplitudes, $m=0, \pm 1$.  Of those the
equal spin pair amplitudes, $m=\pm 1$, lead to the long-range
proximity effect, as pairing occurs within the same spin band and is
unaffected by the large exchange energy $J$. This explains the
penetration of triplet pair amplitudes into the ferromagnet, once they
are created \cite{bergeret01,bergeret05,kadigrobov01}. However, the
questions how they are created in the first place and what are their
magnitudes remain active research topics
\cite{bergeret01,bergeret05,kadigrobov01,eschrig03,lofwander05,champel05,heikkila00,Fom2003,buzdin05,nazarov_condmat,tanaka_condmat}.

The creation mechanisms for pairing correlations inside the
ferromagnet differ in the ballistic limit significantly for weak and
strong ferromagnets. For the following discussion we will concentrate
on singlet superconductors. In the
case of a weak ferromagnet the singlet superconductor induces in the
first place a mixture of singlet and $m=0$ triplet pairs in the
ferromagnet. Both penetrate only on the short magnetic length scale,
but various mechanisms may create long-range $m=\pm 1$ triplet
components. Examples are (i) a magnetic 
domain wall near the interface within a distance of the order of the
magnetic length;\cite{bergeret01,konstandin05} or (ii) two
ferromagnets with misaligned quantization axes separated by a singlet
superconductor with a thickness of the order of the superconducting
coherence length;\cite{Fom2003,lofwander05} or (iii) a spin-active
interface that allows for spin-flip processes.\cite{eschrig03} In all
cases, magnetic inhomogeneities mix the triplet pair components and
create the long-range equal-spin pair amplitudes.

In the case of a strong ferromagnet the roles of the ferromagnet
and the superconductor are reversed. Here, in the first place the
ferromagnet acts as a source for spin-polarization of Cooper pairs
in the superconductor. This results in a boundary layer with
coexisting singlet and triplet amplitudes near the interface
extending about a coherence length into the superconductor.  The
important mechanisms here are the spin-mixing
terms\cite{eschrig03,tok88,fog00,bar02,ZLS_spin,cot04} (often called
spin-rotation) in the reflection and transmission amplitudes of the
surface scattering matrix.  Such spin-mixing effects arise as a
consequence of the different matching conditions for spin-up and
spin-down wave functions at the interface.\cite{tok88} Consequently,
the creation of triplet pair amplitudes is entirely the result of the
interface properties, taking place in an interface region that is of
similar size as the magnetic length. Spin-mixing is most effective at
interfaces with strong ferromagnets, increasing in strength with growing
spin-polarization of the ferromagnet.  Long-range triplet components
are created when spin-flip centers are present in the interface
region.  This mechanism even works in the presence of completely
polarized ferromagnets, since the triplet correlations are created
entirely within the superconductor, and only after their creation
penetrate into the ferromagnet.\cite{eschrig03} In addition, the
magnitude of the triplet correlations at the interface is proportional
to that of the singlet amplitude at the interface, and both are
insensitive to impurity scattering (in contrast to the decay behavior
away from the interface).\cite{cherno}

For intermediate spin polarizations the two creation mechanisms for
triplet pairing, with strengths depending on microscopic details of
the interface and domain wall structures in the ferromagnet, compete
and are difficult to characterize theoretically. However, the two
limiting cases of weak and strong spin polarization can be treated
within a controlled approximation, namely the quasiclassical theory of
superconductivity.\cite{eil,lar} It relies on a separation of two
energy scales, the superconducting gap $\Delta$ and the Fermi energy
$E_f$, and can be extended to superconductor-ferromagnet
heterostructures when the exchange energy $J$ lies either in the low-
or in the high-energy range, corresponding to weak and strong
ferromagnets, respectively. For weak ferromagnets, this means that
Fermi surface properties like the density of states at the Fermi level
in the normal state, $N_f$, and the Fermi velocity, $v_f$, remain
unchanged to lowest order in the small quasiclassical expansion
parameter $J/E_f$; note that the magnetization in the normal state is
a first order term in this parameter and is given by $2\mu_B N_f J$,
where $\mu_B$ is the Bohr magneton.  On the other hand, for strong
ferromagnets all effective interactions and indeed the quasiparticle
band structure itself will be strongly modified by the presence of the
exchange splitting. It is not possible a priory to describe the
crossover between the two limits within quasiclassical theory.

The outline of this article is as follows. In Sec.~2 we describe
how the Pauli principle leads naturally to a classification of
superconducting correlations according to their symmetries with
respect to spin, momentum, and energy. We then show by two examples
that all these types of correlations are indeed induced at
superconductor-ferromagnet interfaces.  In Sec.~3.1 we analyze the
case of a weak ferromagnet in contact with a singlet superconductor
through a spin-active interface barrier. We also illustrate the
difference between the spatial dependences of short-range and
long-range proximity amplitudes. In Sec.~3.2 we consider the case
of a strong ferromagnet.  In particular, we show results for the
proximity effect between a singlet superconductor and a half-metal,
and for a Josephson junction involving a half metal. We show that an
analytic treatment is possible for the case of small transmissions and
a small spin-mixing angle. We find that in these limits, a previously
discovered\cite{eschrig03} low-temperature anomaly in the critical
Josephson current is independent of the interface parameters, and is a
direct consequence of the different symmetry properties of the pairing
correlations compared to the case of a normal metal between two
superconductors.

\section{SYMMETRY CLASSIFICATIONS}

There has been considerable work that formulated the physics of
pairing correlations in diffusive ferromagnets in terms of one
particular pairing component, that has odd-frequency, $s$-wave,
equal-spin triplet symmetry.\cite{bergeret05} The question arises to
which extend this component is important for systems with weak or
intermediate impurity scattering.  We will show in the following
chapters that in fact four different symmetry components exist in
ballistic heterostructures, and that they are comparable in size. We
start with a general classification of pairing correlations.

Superconducting correlations are quantified by the anomalous Green's
function
\begin{equation}
{\cal F}_{\alpha\beta}({\vec r}_1,\tau_1;{\vec r}_2,\tau_2) =
\left< T_{\tau} \Psi_{\alpha}(\vec{r}_1,\tau_1)\Psi_{\beta}(\vec{r}_2,\tau_2) \right>.
\label{agf}
\end{equation}
It is a matrix in spin space and depends on two coordinates and, in
the Matsubara technique, on two imaginary times.  The Pauli principle
requires that this function changes sign when the two particles are
interchanged,
\begin{equation}
{\cal F}_{\alpha\beta}({\vec r}_1,\tau_1;{\vec r}_2,\tau_2) =
- {\cal F}_{\beta\alpha}({\vec r}_2,\tau_2;{\vec r}_1,\tau_1).
\end{equation}
This well-known condition follows directly from Eq.~(\ref{agf}) and
the anti-commutation relations for the field operators.
For homogeneous systems $\cal F$ depends only on relative coordinates
${\vec r}={\vec r}_1-{\vec r}_2$ and $\tau=\tau_1-\tau_2$, i.e. it
follows ${\cal F}_{\alpha\beta}({\vec r},\tau) = -{\cal
F}_{\beta\alpha}(-{\vec r},-\tau)$, and after Fourier transformations
\begin{equation}
{\cal F}_{\alpha\beta}({\vec p},\epsilon_n) = -{\cal F}_{\beta\alpha}(-{\vec p},-\epsilon_n).
\label{Fsymm}
\end{equation}
For inhomogeneous systems this equation holds for each set of center
coordinates. The symmetry restriction Eq.~(\ref{Fsymm}) in spin,
momentum ${\vec p}$, and Matsubara frequency, $\epsilon_n=(2n+1)\pi
T$, can be satisfied in four different ways, listed in
Table~\ref{table_symmetry}. Analytical continuation to the complex
$z$-plane leads to ${\cal F}_{\alpha\beta}({\vec p},z) = -{\cal
F}_{\beta\alpha}(-{\vec p},-z)$; in particular, retarded and advanced
functions are related as ${\cal F}^R_{\alpha\beta}({\vec p},\epsilon)
= -{\cal F}^A_{\beta\alpha}(-{\vec p},-\epsilon)$.\cite{serene}

The spin part of Eq.~(\ref{Fsymm}) can be divided up into singlet and
triplet sectors
\begin{equation}
{\cal F}_{\alpha\beta}(\vec{p},\epsilon_n) =
{\cal F}_s(\vec{p},\epsilon_n) \left( i\sigma_y \right)_{\alpha\beta}
+ \vec{\cal F}_t(\vec{p},\epsilon_n)\cdot \left(\vec{\sigma}\, i\sigma_y\right)_{\alpha\beta},
\end{equation}
where $\vec{\sigma}=(\sigma_x,\sigma_y,\sigma_z)$ is the vector
of the three Pauli matrices. The singlet spin
matrix $(i\sigma_y)_{\alpha\beta}$ is odd under the interchange
$\alpha\leftrightarrow\beta$, while the three triplet matrices
$(\vec{\sigma}\, i\sigma_y)_{\alpha\beta}$ are even.  Some insight
about the symmetries in the momentum- and frequency-domains can be
gained by considering the equal-time
correlator.\cite{ber74,bal92,abr95} E.g., for the spin-triplet case
the Pauli principle imposes that the following sum must vanish
\begin{equation}
\vec{F}_t(\vec{r}=0,\tau=0) = T\sum_{\epsilon_n}\sum_{\vec{p}} \vec{\cal F}_t(\vec{p},\epsilon_n) = 0 .
\end{equation}
When the orbital part is odd the electrons avoid each other in real
space, the equal-time correlator $\vec{F}_t(\vec{p},\tau=0)=
T\sum_{\epsilon_n} \vec{\cal F}_t(\vec{p},\epsilon_n)$ can be finite,
and the correlator $\vec{\cal F}_t(\vec{p},\epsilon_n)$ is even in
frequency. On the other hand, when the orbital part is even, the
correlator $\vec{\cal F}_t(\vec{p},\epsilon_n)$ is odd in frequency,
and electrons avoid each other in time.

\begin{table}[t]
\caption{\small The Pauli principle, requiring the pair correlation
function to be an odd function under the exchange of two electrons,
can be met by properties in spin, orbital, or frequency space as
follows:}
\label{table_symmetry}
\begin{center}
  \begin{tabular}{|l||c|c|c||c|}
    \hline
    type & spin           & momentum & frequency & overall symmetry\\
    \hline\hline
       A & singlet (odd)  & even     & even      & odd\\
    \hline
       B & singlet (odd)  & odd      & odd       & odd\\
    \hline
       C & triplet (even) & odd      & even      & odd\\
    \hline
       D & triplet (even) & even     & odd       & odd\\
    \hline
  \end{tabular}
\end{center}
\end{table}

We summarize the symmetry classes in Table~\ref{table_symmetry}. 
The usual spin singlet s-wave orbital symmetry in a BCS
superconductor is of type A, while the spin triplet p-wave orbital
symmetry superfluid formed in $^3$He is of type C. 
Type D was first considered by
Berezinskii \cite{ber74} in connection with early research on
superfluid $^3$He. Finally, type B was considered in connection with
unconventional superconductors by Balatsky, Abrahams and others
\cite{bal92,abr95,coleman93,fuseya03}.

So far we have only considered the correlation function. In order to
obtain the gap function, additional knowledge of the pairing
interaction
$\lambda(\vec{p},\vec{p}^{\;\prime},\epsilon_n,\epsilon_n^{\prime})$
is required (here for the singlet case)
\begin{equation}
  \Delta_s(\vec{p},\epsilon_n) = T\sum_{\epsilon_n^{\prime}}\sum_{p^{\prime}}
  \lambda_s(\vec{p},\vec{p}^{\;\prime},\epsilon_n,\epsilon_n^{\prime})
	 {\cal F}_s(\vec{p}^{\;\prime},\epsilon_n^{\prime}).
\end{equation}
Clearly, the symmetry of the pairing interaction dictates the
symmetry of the gap function through the projection of the correlation
function in the gap equation. In particular, if the superconducting
correlations are odd in frequency a frequency-dependent pairing
interaction (due to strong retardation effects) is needed to obtain a
non-vanishing gap and a superconducting transition.

\section{TRIPLET PAIRING IN CLEAN S/F STRUCTURES}

\subsection{Weak Ferromagnets}

To illustrate how triplet correlations are induced we study a
superconductor-weak ferromagnet heterostructure in the ballistic
transport regime and include spin-active interface scattering. For
simplicity we consider temperatures near the superconducting critical
temperature; however, the main results of this section require only
the pairing amplitudes in the ferromagnet to be small, and apply with
minor modifications to any temperature.  In this case, within
quasiclassical approximation, the anomalous Green's function follows
from the (linearized) Eilenberger equations,\cite{eil,lar}
\begin{eqnarray}
\left( \vec{v}_f\cdot\mbox{\boldmath$\nabla$}
 + 2\epsilon_n \right) f_s &=& 2\pi\Delta\, \mbox{sgn}(\epsilon_n) - 2iJ\, f_{tz},\label{deq1}\\
\left( \vec{v}_f\cdot\mbox{\boldmath$\nabla$}
 + 2\epsilon_n \right) f_{tz} &=& -2iJf_s,\label{deq2}\\
\left( \vec{v}_f\cdot\mbox{\boldmath$\nabla$}
 + 2\epsilon_n \right) \vec{f}_{t\perp} &=& 0.\label{deq3}
\end{eqnarray}
Here, $v_f$ denotes the Fermi velocity for the quasiparticles in the
respective material.  The superconducting gap $\Delta$ is non-zero
only in the superconductor, while the exchange field $J$ is non-zero
only in the ferromagnet. We assume that the ferromagnet has a single
domain and use the direction of its exchange field $J\hat z$ as spin
quantization axis. The spin-active interface can have a different
spin-quantization axis, that we denote $\hat\mu$ (see below).

The main features of the set of differential equations
(\ref{deq1})-(\ref{deq3}) are:
(i) The inhomogeneity of the equation for $f_{tz}$ requires both $J$
and $f_s$ to be present.  Thus, this component naturally emerges in a
ferromagnet coupled to a singlet superconductor.
(ii) The eigenvalues of the $f_s$-$f_{tz}$ sub-system for a
particular $\vec{v}_f$ are given by $k_n^{\pm}=2(|\epsilon_n|\pm
iJ)/v_f$.  Thus, both the singlet $f_s$ and the triplet $f_{tz}$
oscillate on the clean-limit magnetic length scale
$\xi_J=v_f/2J$, and decay exponentially on the length scale
$\xi_n=v_f/2|\epsilon_n|$; the latter is dominated by the lowest
Matsubara frequency, $\epsilon_0=\pi T$, and occurs on the clean-limit
normal-metal coherence length scale $\xi_T=v_f/2\pi T$.
(iii) The equation for ${\vec f}_{t\perp}=(f_{tx},f_{ty})$ is
decoupled from the others and is homogeneous. Therefore, the presence
of these components requires spin-active interface scattering.
(iv) The equations for $\vec{f}_{t\perp}$ do not contain the exchange
field, and these components are monotonic decaying functions on the
scale $\xi_n$.

\begin{figure}[t]
\includegraphics[width=\textwidth]{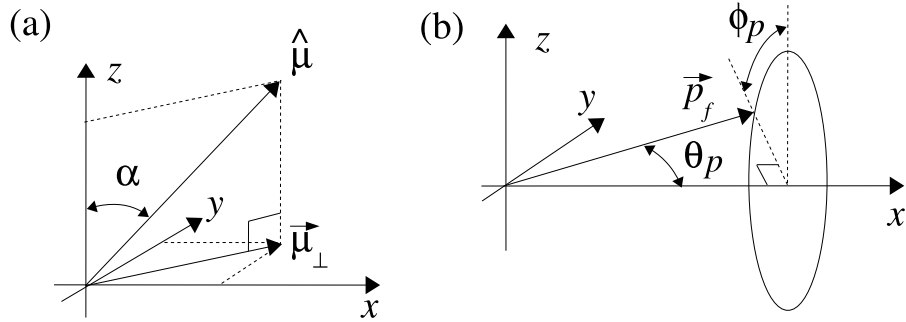}
\caption{\small (a) The spin quantization axis $\hat\mu$ of the
interface spin-active scattering matrix forms an angle $\alpha$ with
the exchange field in the ferromagnet and has a perpendicular
component $\vec{\mu}_{\perp}$. (b) The direction of the momentum
$\vec{p}_f$ is expressed in spherical coordinates through the angles
$\theta_p$ and $\phi_p$.}
\label{fig_geometry}
\end{figure}

The Eilenberger equations are solved by integrating along trajectories
$\vec{v}_f(\vec{p}_f)$ with an initial condition at the starting point
of the trajectory. We consider the three-dimensional case and
introduce a spherical coordinate system as shown in
Fig.~\ref{fig_geometry} (b).  The superconductor occupies the region $x<0$,
and the ferromagnet the region $x>0$.  We assume rotational invariance
around $\hat x$, i.e. all quantities will be independent of the
variable $\phi_p$.

For positive Matsubara frequencies, the initial condition for the
trajectories $\theta_p\in[0,\pi/2]$ deep in the superconductor is
$f(x\rightarrow -\infty)=(\pi\Delta/|\epsilon_n|)i\sigma_y$. For
trajectories $\theta_p\in[\pi/2,\pi]$ we start the integration deep in
the ferromagnet with initial condition $f(x\rightarrow +\infty)=0$. For
negative Matsubara frequencies, the stable direction of integration is
the opposite. After applying the boundary conditions described below,
the integration proceeds away from the interface.

The boundary conditions at the superconductor-ferromagnet interface
for the 2x2 spin matrix Green's functions coincide for positive
Matsubara frequencies with those for the retarded
functions\cite{ZLS_spin}. Near $T_c$, they read
\begin{equation}
f_{2,out} = S_{22} f_{2,in} \underline S_{22}^{\dagger} + S_{21} f_{1,in} \underline S_{21}^{\dagger},
\label{f_bc}
\end{equation}
where the indices $1$ and $2$ refer to the superconducting and
ferromagnetic sides, respectively, and `$in$' and `$out$' denote
functions with momenta directed towards or away from the
interface. The scattering matrix for holes is written in terms of the
scattering matrix for electrons as $\underline
S_{ij}(\vec{p}_{\parallel})=S_{ji}(-\vec{p}_{\parallel})^T$. For
negative Matsubara frequencies the boundary conditions coincide with
the ones for advanced functions, which here simply means interchanging
$S\leftrightarrow S^{\dagger}$ and the side indices ($i\leftrightarrow
j$).

For definiteness, we consider a simplified spin-active model in which
the scattering matrix is independent of the sign of the parallel
momentum $\vec{p}_{\parallel}$. In this case e.g. the scattering
matrix for transmission can be written as\cite{ZLS_spin}
\begin{equation}
S_{12}=S_{21} = e^{i\varphi/2}\left[ s_{21}+s^{\prime}_{21}(\hat\mu\cdot\vec{\sigma})\right]
\exp\left[i(\hat\mu\cdot\vec{\sigma})\vartheta/2\right],
\end{equation}
where we have an average transmission parameter
$s_{21}=(|t_+|+|t_-|)/2$, a spin-filtering transmission parameter
$s^{\prime}_{21}=(|t_+|-|t_-|)/2$, plus a spin-mixing angle
$\vartheta$. The amplitudes $t_{\pm}$ denote the transmission
amplitudes of the interface for the two spin-directions in the basis
where $\hat\mu\cdot\vec{\sigma}$ is diagonal. The phase $\varphi$ does
not play a role in the following.

For small transmission and small spin mixing the energy gap $\Delta$
has a step function form, in which case $f_{1,in}$ retains its bulk
form at the interface. Similarly, the incoming function on the F-side
retains its bulk form, i.e. it vanishes $f_{2,in}=0$. The resulting
outgoing amplitude at the interface on the ferromagnetic side can be
written as $f_{2,out} = (A_s + \mbox{sgn}(\epsilon_n)A_t\,
\hat\mu\cdot\vec\sigma)i\sigma_y$, where
\begin{equation}
A_s = A_0\cos \vartheta,\hspace{1cm}
A_t = iA_0\sin \vartheta,
\end{equation}
and the prefactor is $A_0=|t_+|\,|t_-|\,\pi\Delta/|\epsilon_n|$. These
amplitudes are the initial conditions for the singlet and triplet
components in the ferromagnet at the interface. The spatial dependence
of all amplitudes can be expressed in terms of the effective
coordinate $x_\theta = x/|\cos \theta_p |$.  We obtain the following
explicit expressions in the ferromagnet
\begin{eqnarray}
f_s(\epsilon_n,x_\theta) &=& c_s
                    \left[ A_s \cos\left(\frac{x_\theta}{\xi_J}\right) 
                          - iA_t\cos\alpha \, \sin\left(\frac{x_\theta}{\xi_J}\right) \right]
                    e^{-k_nx_\theta},\\
f_{tz}(\epsilon_n,x_\theta) &=& c_s\,\mbox{sgn}(\epsilon_n)
                    \left[ A_t\cos\alpha \,  \cos\left(\frac{x_\theta}{\xi_J}\right) 
                          - iA_s\sin\left(\frac{x_\theta}{\xi_J}\right) \right]
                    e^{-k_nx_\theta},\quad\quad\\
{\vec f}_{t\perp}(\epsilon_n,x_\theta) &=& c_s\,\mbox{sgn}(\epsilon_n)\,
                                A_t e^{-k_nx_\theta}\,{\vec \mu}_{\perp},
\end{eqnarray}
where $k_n=2|\epsilon_n|/v_f$.  Here we introduced the components of
$\hat\mu$ parallel to the z-axis, $\cos\alpha \, \hat z$, and
perpendicular, ${\vec\mu}_{\perp}$, see Fig.~\ref{fig_geometry} (a). The
coefficient $c_s=\left[1+\mbox{sgn}(\vec{p}_f\cdot\hat x)\,
\mbox{sgn}(\epsilon_n)\right]/2$ selects the correct sign of the
momentum relative to the x-axis.

We conclude at this stage:
(i) The singlet component is purely real while the triplet components
are purely imaginary. It follows that in the complex plane
$f_s(-z^\ast )=f_s(z)^\ast $, $\vec{f}_t(-z^\ast )=-\vec{f}_t(z)^\ast$
for each $\vec{p}$.  E.g., the retarded functions have the symmetries
$f^R_s(-\epsilon )=f^R_s(\epsilon )^\ast $ and $\vec{f}^R_t(-\epsilon
) =-\vec{f}^R_t(\epsilon)^\ast$.
(ii) Spin-mixing (finite $\vartheta$) is crucial for triplet
components to be induced at the interface.
(iii) The components ${\vec f}_{t\perp}$ are induced only when
$\hat\mu$ is misaligned with respect to the exchange field (here
$J\hat z$).

To investigate the symmetry properties we expand the correlation
functions in partial waves. Since we have rotational invariance around
the $\hat{x}$-axis, and the correlation functions only depend on
$\cos\theta_p$ through the projection of the Fermi velocity on the
x-axis, we can expand in Legendre polynomials $P_l(\cos \theta_p)$ and
get
\begin{eqnarray}
f_s(\cos\theta_p,\epsilon_n,x) &=& \sum_{l=0}^{\infty} f_s(l,\epsilon_n,x) P_l(\cos\theta_p),\\
f_s(l,\epsilon_n,x) &=& \frac{2l+1}{2} \int_{-1}^1 d(\cos\theta_p) f_s(\cos\theta_p,\epsilon_n,x) P_l(\cos\theta_p),
\label{Legendre-expansion}
\end{eqnarray}
and similarly for the triplet components. To proceed we have to model
the dependencies of the tunneling probabilities and the spin-mixing
angle on the trajectory angle and then evaluate the various
harmonics. The simplest model is a tunnel cone model with constant
transmission probabilities and spin-mixing angle within a range of
trajectory angles, i.e.  $|t_+(\theta_p)| = \sqrt{{\cal T}_+}$,
$|t_-(\theta_p)| = \sqrt{{\cal T}_-}$, and $\vartheta(\theta_p)$
constant simply called $\vartheta$.  In the following we assume a very
wide tunnel cone ($\rightarrow\pi/2$), but it is straightforward to
obtain the expressions for a more narrow cone. After integrations over
$\cos\theta_p$ we obtain the amplitudes
\begin{eqnarray}
f_s(l) &=& A_0\left[\mbox{sgn}(\epsilon_n)\right]^l
\left[\cos\vartheta \; {\rm Re} Q_l(z) - \cos\alpha\sin\vartheta \;{\rm Im} Q_l(z)\right],\label{fsSA}\\
f_{tz}(l) &=& iA_0 \left[\mbox{sgn}(\epsilon_n)\right]^{l+1}
\left[\cos\alpha \sin\vartheta \; {\rm Re} Q_l(z)+\cos\vartheta \; {\rm Im} Q_l(z) \right],
\;\;\;\;\label{ftpSA1}\\
{\vec f}_{t\perp}(l) &=& iA_0 \left[\mbox{sgn}(\epsilon_n)\right]^{l+1}
\left[\sin\vartheta\; Q_l(k_nx)\right] {\vec\mu}_{\perp},\label{ftpSA}
\end{eqnarray}
where $z=k_n^{+}x=2(|\epsilon_n|+iJ)x/v_f$.  The function $Q_l(z)$ for
the first few $l$ is given by $Q_0(z) = z\Gamma(-1,z)/2$ ($s$-wave),
$Q_1(z) = 3z^2\Gamma(-2,z)/2$ ($p$-wave), and $Q_2(z) = (5/2)\left[
3z^3\Gamma(-3,z)/2 - z\Gamma(-1,z)/2\right]$ ($d$-wave), where
$\Gamma(n,z)$ is the upper incomplete gamma-function. Note the
different spatial dependences of the singlet and $m=0$ triplet on the
one hand, and the perpendicular triplets on the other hand, as they
enter in the argument of $Q_l$. Also note that $Q_l(k_nx)$ is purely
real since $k_nx$ is real. The Pauli principle and the symmetry
requirements it imposes is affirmed in
Eqs.~(\ref{fsSA})-(\ref{ftpSA}). We can also conclude that we have all
four types of correlations, see Table~\ref{table_weak},
including equal-spin pairing correlations when $\hat\mu$ is not
parallel to $\vec{J}$.
\begin{table}[t]
\caption{\small The pairing correlations in ballistic
superconductor-ferromagnet systems with spin-active interface
scattering can be identified with the four types listed in
Table~\ref{table_symmetry}.}
\label{table_weak}
\begin{center}
  \begin{tabular}{|l|l|c|}
    \hline
    Eq.~(\ref{fsSA}), singlet: & even $l$: even parity, even in frequency & type A\\
    \cline{2-3}
                               &  odd $l$:  odd parity,  odd in frequency & type B\\
    \hline
    Eqs.~(\ref{ftpSA1})-(\ref{ftpSA}), triplets: & odd $l$: odd parity, even frequency & type C\\
    \cline{2-3}
                                                & even $l$: even parity, odd frequency & type D\\
    \hline
  \end{tabular}
\end{center}
\end{table}

\begin{figure}[t]
\centerline{\includegraphics[width=0.8\textwidth]{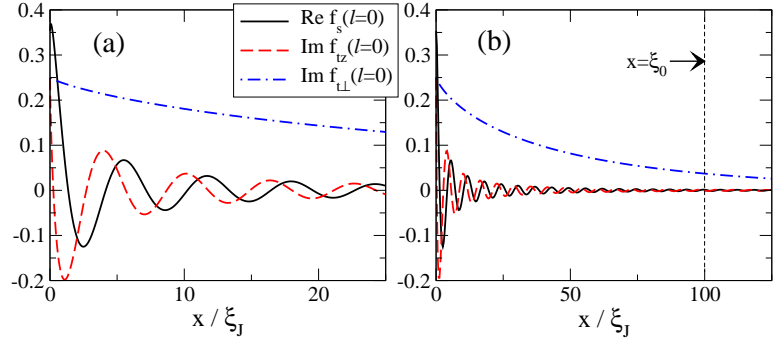}}
\caption{\small Pair correlation functions in the ferromagnet of a
ballistic superconductor-ferromagnet junction. The first partial wave
($l=0$) for the lowest Matsubara frequency, $\epsilon_0=\pi T$, is
shown. (a) For the singlet, $f_s$, and triplet with zero
spin-projection on the exchange field, $f_{tz}$, we observe a
$1/x$-decay and oscillations, both on the scale $\xi_J=v_f/2J$. On the
same scale the perpendicular triplet $f_{t\perp}$ only varies slowly.
(b) At large distances all components decay on the coherence length
scale $\xi_0$. However, the magnitudes of $f_s$ and $f_{tz}$ are
considerably reduced compared with $f_{t\perp}$ before this region is
reached. We have chosen $\xi_0=100\xi_J$, $\alpha=\pi/4$,
$\vartheta=\pi/4$, and we have normalized the functions by
$(\Delta/T)\sqrt{{\cal T}_+{\cal T}_-}$.}
\label{fig_triplet}
\end{figure}

In the region $\xi_J\ll x \ll \xi_0$ of the ferromagnet the various
components show different decaying behaviors depending on whether the
pairing correlations involve two spin bands ($f_s$ and $f_{tz}$) or
only one spin band ($\vec{f}_{t\perp}$).  The asymptotic form of the
gamma function for $|z|\gg 1$ is $\Gamma(n,z)\sim z^{n-1}e^{-z}$, and
we can obtain simplified expressions for $x\gg\xi_J$. We focus on the
lowest Matsubara frequency, for which
\begin{equation}
\left.Q_l(z)\right|_{x\gg\xi_J} \approx
-i\,\frac{2l+1}{2}\frac{\xi_J}{x}e^{-x/\xi_0}e^{-ix/\xi_J},
\end{equation}
where $\xi_0=v_f/2\pi T_c$. The singlet and triplet amplitudes with
zero spin-projection have the forms
\begin{eqnarray}
f_s(l) &=& f_0
\left[ -\cos\vartheta \, \frac{\sin\left(x/\xi_J\right)}{x/\xi_J} 
       + \cos\alpha\, \sin\vartheta \, \frac{\cos\left(x/\xi_J\right)}{x/\xi_J} \right]
e^{-x/\xi_0},\label{asympt1}\\
f_{tz}(l) &=& if_0
\left[ -\cos
\vartheta \, \frac{\cos\left(x/\xi_J\right)}{x/\xi_J}
       - \cos\alpha \sin\vartheta \, \frac{\sin\left(x/\xi_J\right)}{x/\xi_J} \right]
e^{-x/\xi_0},\label{asympt2}
\end{eqnarray}
where the prefactor is $f_0=\sqrt{{\cal T}_+{\cal
T}_-}(\Delta/T)(2l+1)/2$.  Note that there is a $\pi/2$ phase shift
between the oscillations of the two components $f_s$ and $f_{tz}$.
Also note that the trajectory resolved functions decay exponentially
on the large coherence length scale $\xi_0$, while the various
harmonics decay as $(x/\xi_J)^{-1}$, before the exponential decay on
the $\xi_0$ scale sets in at large distances; the difference comes
from the average over trajectories when projecting out the various
harmonics.  On the other hand, for the perpendicular triplets the
asymptotic behavior is not reached until $x\gg\xi_0$, and is of the
form
\begin{equation}
{\vec f}_{t\perp}(l) \approx if_0
\sin\vartheta\, \frac{\xi_0}{x}
e^{-x/\xi_0} {\vec\mu}_{\perp} .
\label{asympt3}
\end{equation}
We plot in Fig.~\ref{fig_triplet} the $l=0$ component of the Green's
functions in the ferromagnet, as given by
Eqs.~(\ref{fsSA})-(\ref{ftpSA}) for the lowest Matsubara
frequency. The higher order partial waves look very similar and have
similar amplitudes. As can be seen in the figure and from
Eqs.~(\ref{asympt1})-(\ref{asympt2}), in the region $\xi_J\ll x \ll
\xi_0$ the correlation functions $f_s$ and $f_{tz}$ decay like $1/x$
and are rapidly reduced by a factor $\xi_J/\xi_0$ compared to
$\vec{f}_{t\perp}$.  A similar decay for $\vec{f}_{t\perp}$ is absent
in that region.  For $x\gg \xi_0$ all components continue to decay
according to $x^{-1}e^{-x/\xi_0}$, as can be inferred from
Eqs.~(\ref{asympt1})-(\ref{asympt3}).

In fact, the Pauli principle requires odd-frequency amplitudes to be
present in any inhomogeneous superconducting state, not necessarily
spin-polarized.  For example, the case of a normal metal coupled to a
superconductor is easily obtained from the above calculations and we
recover for the case of a usual tunnel barrier (no spin-active
scattering) the usual proximity effect \cite{kogan82}. There is only a
singlet component of the Green's function and in the normal metal
region it can be expressed in terms of partial waves as
\begin{equation}
f_s(l) = {\cal T} \frac{\pi\Delta}{|\epsilon_n|}
\left[\mbox{sgn}(\epsilon_n)\right]^l Q_l\left(\frac{2|\epsilon_n|x}{v_f}\right),
\end{equation}
which for large distances from the interface $x\gg\xi_0$ decays as
$\sim (\xi_0/x)\exp(-x/\xi_0)$. Above we introduced ${\cal T}$, the
transparency of the interface within the tunnel cone (here for
simplicity chosen very wide $\rightarrow\pi/2$ as above).  We see that
higher partial waves $l\geq 1$ are induced, as is always the case in
inhomogeneous systems. In accordance with the Pauli principle, all odd
components ($l=1,3,..$) are odd in frequency. Thus, both singlet
entries (types A and B) in Table~\ref{table_symmetry} are always
present in the usual proximity effect, and in fact in any
inhomogeneous singlet superconducting state.  Analogously, the two
triplet entries (types C and D) in Table~\ref{table_symmetry} are
characteristic for any inhomogeneous triplet superconducting state.

\subsection{Half-metallic Ferromagnet}

In this section we analyze the various symmetry components, introduced
in the previous sections, for the case of a superconductor coupled to
a half-metallic ferromagnet. This case was previously studied on
theoretical grounds, and a non-zero Josephson-effect was predicted for
a superconductor/half-metal/superconductor junction.\cite{eschrig03}
The effect has recently been confirmed experimentally.\cite{keizer06}
In our previous article we focused on the $p$-wave triplet amplitudes,
but it should be noted that there is also an odd-frequency $s$-wave
triplet amplitude present in the half-metal. In fact, within our model
all symmetry classes discussed in the previous chapter and shown in
Table~\ref{table_symmetry} are present in the heterostructure. In the
following we discuss the results for the clean case. We have performed
calculations with impurity scattering ranging from the clean limit to
the diffusive limit and have confirmed that both the even-frequency
$p$-wave triplet and the odd-frequency $s$-wave triplet components are
always present and important for the Josephson current through the
half metal.\cite{cherno}

The full scattering matrix for our model is a 3x3 matrix, where the
three scattering channels are the two spin channels in the
superconductor and one metallic spin channel in the half metal. The
second spin channel in the half metal is insulating and is not
participating actively in transport phenomena. As discussed in the
introduction, quasiclassical theory is therefore applicable. To be
specific, we parameterize the scattering matrix connecting incoming to
outgoing waves in the superconductor ($S$) and half-metallic ($F$) sides as,\cite{eschrig03}
\begin{equation}
\label{scatt1}
\hat {\bf S}= 
\left( \begin{array}{c|c}
r^{S}&t^{SF} \\
\hline
t^{FS}&-r^{F}
\end{array}
\right)=
\left( \begin{array}{cc|c}
r_{\uparrow\uparrow}e^{\frac{i}{2}\vartheta} & 
r_{\uparrow\downarrow}e^{i(\vartheta_{\uparrow\uparrow}-\vartheta_{\downarrow\uparrow})}& 
d_{\uparrow\uparrow}e^{i(\vartheta_{\uparrow\uparrow}+\frac{\vartheta }{4})}\\
r_{\downarrow\uparrow}e^{-i(\vartheta_{\uparrow\uparrow}-\vartheta_{\downarrow\uparrow})}&
r_{\downarrow\downarrow}e^{-\frac{i}{2}\vartheta} & 
d_{\downarrow\uparrow}e^{i(\vartheta_{\downarrow\uparrow}-\frac{\vartheta }{4})}\\
\hline
d_{\uparrow\uparrow}e^{-i(\vartheta_{\uparrow\uparrow}-\frac{\vartheta }{4})}&
d_{\downarrow\uparrow}e^{-i(\vartheta_{\downarrow\uparrow}+\frac{\vartheta }{4})}
& - \underline{r}_{\uparrow\uparrow}
\end{array}
\right),
\end{equation}
where the extra phases of the transmission amplitudes 
ensure the unitarity of the scattering matrix,
and the magnitudes are given by
$r_{\uparrow\uparrow }= 1 -t_{\uparrow\uparrow}^2/2W$,
$r_{\downarrow\downarrow }= 1 -t_{\downarrow\uparrow}^2/2W$,
$r_{\uparrow\downarrow }=r_{\downarrow\uparrow }
=  -t_{\uparrow\uparrow}t_{\downarrow\uparrow}/2W$,
$\underline{r}_{\uparrow\uparrow}=
1-(t_{\uparrow\uparrow}^2+t_{\downarrow\uparrow}^2)/2W$,
$d_{\uparrow\uparrow }=t_{\uparrow\uparrow }/W$, and
$d_{\downarrow\uparrow }=t_{\downarrow\uparrow }/W$, 
with $W=1+(t_{\uparrow\uparrow}^2+t_{\downarrow\uparrow}^2)/4$.
The five real parameters of the scattering matrix above
are $t_{\uparrow\uparrow}$, $t_{\downarrow\uparrow}$,
$\vartheta $, $\vartheta_{\uparrow\uparrow}$ and $\vartheta_{\downarrow\uparrow}$.

In the following we derive analytic
expressions for a well defined limiting case: the case of weak
transmissions in all channels and small spin mixing angle. In the
above scattering matrix this means that $\vartheta $,
$t_{\uparrow\uparrow}$ and $t_{\downarrow\uparrow}$ are small.
Thus, in this limit the scattering matrix assumes the form,
\begin{equation}
\label{scatt}
\hat {\bf S}_{0}=
\left( \begin{array}{cc|c}
e^{\frac{i}{2}\vartheta} & 0& t_{\uparrow\uparrow}e^{i(\vartheta_{\uparrow\uparrow}+\frac{\vartheta }{4})}\\
0&e^{-\frac{i}{2}\vartheta} & t_{\downarrow\uparrow}e^{i(\vartheta_{\downarrow\uparrow}-\frac{\vartheta }{4})}\\
\hline
t_{\uparrow\uparrow}e^{-i(\vartheta_{\uparrow\uparrow}-\frac{\vartheta }{4})}&
t_{\downarrow\uparrow}e^{-i(\vartheta_{\downarrow\uparrow}+\frac{\vartheta }{4})}
& - 1
\end{array}
\right).
\end{equation}

Before presenting the results, we comment on the parameters in the
scattering matrix. Within our theory, the scattering matrix is a
phenomenological input that characterizes the scattering of
quasiparticles near the Fermi surface on either side of the interface
of the system.  Microscopic details of the interfaces are irrelevant
for the low-energy physics near the Fermi surfaces. Thus, all the
parameters are assumed to be independent of energy, and only depend on
the positions of the scattering momenta on the Fermi surface. For
strong ferromagnets, $J$ is not a small quantity, and consequently the
scattering matrix has spin-active scattering terms of the order of
$J/\epsilon_F$, with $\epsilon_F$ the Fermi energy. Whereas the
spin-mixing angle $\vartheta $ is a robust property of any interface
to a strong ferromagnet, the spin-flip parameters
$t_{\downarrow\uparrow }$, $\vartheta_{\downarrow\uparrow}$ require
the breaking of spin-rotation symmetry around the quantization axis of
the ferromagnet. This can be the result of various mechanisms, for
example the presence of interface regions with
misaligned spin (magnetic grain layers).  The above mentioned parameters 
represent in this case averages over the grain configuration along the
contact region of the interface for {each given sample},
$t_{\downarrow\uparrow }e^{i\vartheta_{\downarrow\uparrow}}=
\big\langle t^{(i)}_{\downarrow\uparrow }
e^{i\vartheta^{(i)}_{\downarrow\uparrow}} \big\rangle_{i} $, where $i$
numbers the different grains. Thus, there are expected
sample-to-sample variations in the spin-flip parameters unless the
typical grain size exceeds the contact size.  

We now proceed with the derivation of analytic expressions for the
various symmetry components of the pairing amplitudes.  On the
superconducting side of an interface with a half metal, all four
components listed in Table \ref{table_symmetry} are induced and can be
calculated perturbatively in the small
parameters $\vartheta $, $t_{\uparrow\uparrow}$ and
$t_{\downarrow\uparrow}$.  In the following we denote
$\Omega_n=\sqrt{\epsilon_n^2+|\Delta|^2}$.  We obtain for type D:
$f_{tz}^D=i\pi \vartheta \epsilon_n \Delta /2\Omega_n^2 $, for type C:
$f_{tz}^C=-i\pi \vartheta \Delta /2 \Omega_n$.  In an expansion in
spherical harmonics the two components $f_{tz}^D$ and $f_{tz}^C$
correspond to $s$-wave and $p$-wave triplets.  The type B component only
enters in second order in $\vartheta $, and has the form $f_{s}^B=\pi
\vartheta^2 \epsilon_n \Delta/4\Omega_n^2 $.  Finally, the
renormalization of the type A singlet component, $\delta f_{s}^A=-\pi
\vartheta^2 \epsilon_n^2 \Delta /4\Omega_n^3$, reduces the leading
component only in second order in $\vartheta$.  Consequently, to
linear order in $\vartheta$ the self consistent singlet order
parameter $\Delta $ is not affected and stays constant up to the
interface.  All components, $\delta f_{s}^A$, $f_{s}^B$, $f_{tz}^C$,
and $f_{tz}^D$ decay into the superconductor exponentially with a
decay length given by $\xi^{(S)}_n=v_f/2\Omega_n$.

Next we consider the equal spin pairing amplitudes in a symmetric
Josephson junction with a half metal of length $L$ extending from
$-L/2<x<L/2$. On either side of the half metal are singlet
superconductors having order parameters $|\Delta |e^{i\chi_1} $ and
$|\Delta |e^{i\chi_2} $.  The equal spin pairing amplitudes in the
half metal, $f_{\uparrow\uparrow}$ 
are proportional to the components $f_{tz}$ in the
superconductors.  For the even-frequency triplet amplitude we obtain
in the half metal,
\begin{equation}
f^C_{\uparrow\uparrow}=-i\; \beta_n \; \mbox{sgn}(\cos \theta_p)
\left( 
\frac{\sinh \frac{k_nx}{\mu }}{\sinh \frac{k_nL}{2\mu }}(e^{i\tilde\chi_2}+e^{i\tilde\chi_1})
+\frac{\cosh \frac{k_nx}{\mu } }{\cosh \frac{k_nL}{2\mu }}(e^{i\tilde\chi_2}-e^{i\tilde\chi_1}) 
\right),
\label{fc}
\end{equation}
and for the odd-frequency triplet amplitude,
\begin{equation}
f^D_{\uparrow\uparrow}=i\; \beta_n\; \mbox{sgn}(\epsilon_n)
\left( 
\frac{\cosh \frac{k_nx}{\mu} }{\sinh \frac{k_nL}{2\mu}}(e^{i\tilde\chi_2}+e^{i\tilde\chi_1}) 
+\frac{\sinh \frac{k_nx}{\mu }}{\cosh \frac{k_nL}{2\mu }}(e^{i\tilde\chi_2}-e^{i\tilde\chi_1})
\right),
\label{fd}
\end{equation}
where $k_n=2|\epsilon_n|/v_f$, $\mu=|\cos \theta_p |$, and $\beta_n=
\pi t_{\uparrow\uparrow} t_{\downarrow\uparrow} \vartheta |\Delta
||\epsilon_n|/ 2(\epsilon_n^2+|\Delta|^2)$.  The superconducting
phases are renormalized by the phase shifts during tunneling, and are
given by $\tilde
\chi_i=\chi_i-\vartheta_{i,\uparrow\uparrow}-\vartheta_{i,\downarrow\uparrow}$
with $i=1,2$.  Note that equal-spin amplitudes are only possible
when both $t_{\downarrow \uparrow} \ne 0 $ and $\vartheta\ne 0$.  Even
and odd frequency (i.e. $p$-wave and $s$-wave) triplets at the
interfaces ($x=\pm L/2$) are comparable in size.

\begin{figure}[t]
\includegraphics[width=\textwidth]{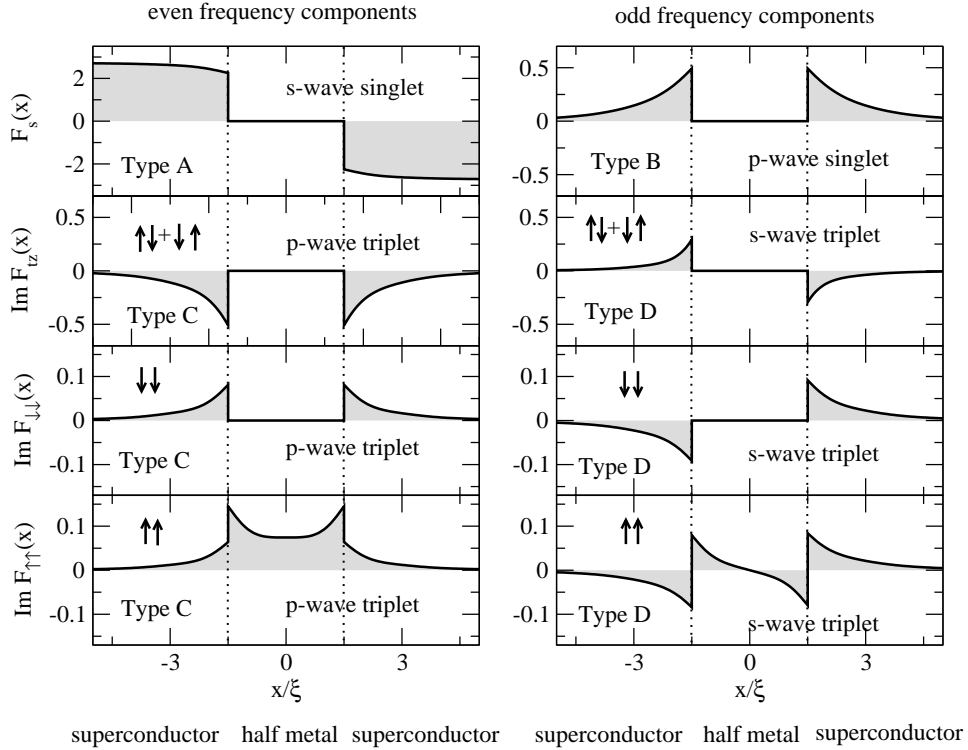}
\caption{\small Spatial dependences of all components of the pairing
amplitude for a symmetric superconductor/half-metal/superconductor
Josephson $\pi$-junction with interfaces characterized by the
scattering matrix Eq.~(\ref{scatt}).  The interfaces are indicated by
the dashed lines.  The components were numerically iterated until self
consistency with the singlet order parameter was achieved.  The
interface parameters for normal impact angle are $\vartheta=0.3\pi $,
$t_{\uparrow\uparrow}=1$, $t_{\downarrow \uparrow}=0.7 $,
$\vartheta_{\uparrow\uparrow}=\vartheta_{\downarrow\uparrow}=0$, and
the temperature is $T=0.2T_c$.  
\label{fig_clean}
}
\end{figure}

\begin{figure}[t]
\includegraphics[width=\textwidth]{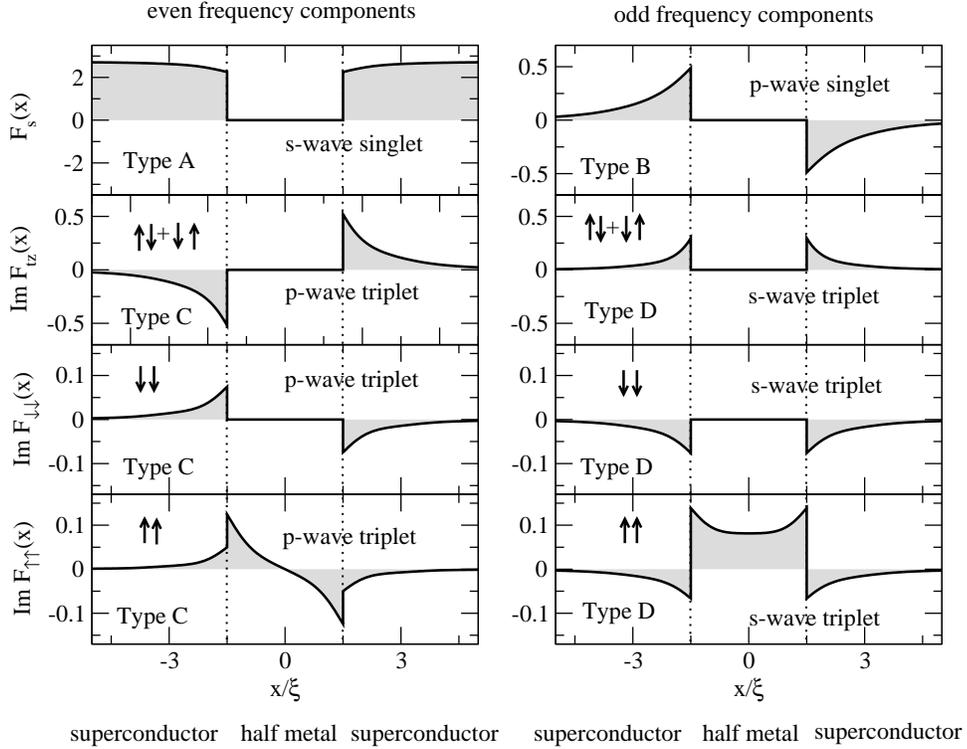}
\caption{\small
The same as in Fig.~\protect\ref{fig_clean}, but for
zero phase difference.
\label{fig_clean_zero}
}
\end{figure}

To illustrate the different symmetry components, we show in
Fig.~\ref{fig_clean} numerical results for a
superconductor/half-metal/superconductor junction.  For these results
we do not expand the interface scattering matrix in $\theta $,
$t_{\uparrow\uparrow } $, $t_{\downarrow\uparrow }$ as in
Eq.~(\ref{scatt}), but use the general scattering matrix 
(\ref{scatt1}) and solve 
the boundary conditions numerically.\cite{eschrig03} We allow for an
angular variation with respect to the interface normal of both the
spin mixing angle and the transmission amplitudes; for definiteness we
assume a variation proportional to $|\cos \theta_p |$. We iterate the
order parameter and the boundary conditions until self consistency is
achieved.  We discuss first results for a $\pi$-junction ($\tilde
\chi_1=0$, $\tilde\chi_2=\pi$) as this is the ground state of the
Josephson junction.\cite{eschrig03} To quantify the spatial
dependences of the different symmetry amplitudes, we define the
functions
\begin{equation} 
F_{\alpha \beta}(x)=
\frac{2}{\pi}T\sum_{\epsilon_n>0} \Big\langle P_l(\cos \theta_p)\;
f_{\alpha \beta}(\epsilon_n,x,\theta_p) \Big\rangle_{\vec{p}} \; f^\ast_h (\epsilon_n),
\end{equation}
where $P_l(\cos \theta_p)$ is chosen to project out the $s$-wave
($l$=0) or $p$-wave ($l$=1) parts of the pairing amplitudes, and
$f_h(\epsilon_n)=\pi\Delta/\Omega_n $ is the propagator for a
homogeneous singlet superconductor that is introduced here to ensure
convergence at large $\epsilon_n$ (note that at the boundary points
$f_{\alpha \beta} (\epsilon_n)$ decays weakly $\sim 1/\epsilon_n$).
As is shown in Fig.~\ref{fig_clean}, on the superconducting side 
all possible symmetry components listed in Table~\ref{table_symmetry}
are induced, of which the even-frequency $p$-wave triplet and the
odd-frequency $s$-wave triplet penetrate into the half metal. The
latter two components are of similar magnitude, however the $p$-wave
triplet is larger in the center region of the junction than the
$s$-wave triplet.  We have also performed calculations including
impurity scattering and have shown that {\it both} components are
essential for the Josephson current in the entire range from the
ballistic to the diffusive limit.  Details will be presented in a
forthcoming publication.\cite{cherno} Also seen in
Fig.~\ref{fig_clean} is that there are different symmetries with
respect to the spatial coordinate, that result from the symmetry of
the Josephson junction. The components shown are for the ground state;
with a finite Josephson current ($\tilde\chi_2-\tilde\chi_1 \ne
0,\pi$) all amplitudes become complex and the spatial symmetries are
lost, as can also be inferred from Eqs.~(\ref{fc})-(\ref{fd}).  Note
that in Fig.~\ref{fig_clean}, as consequence of higher order terms in
the transmission parameters and the spin-mixing angle, there are
equal-spin pairing correlations also in the superconducting regions,
and there are noticeable inhomogeneous contributions to the singlet
amplitudes as well.

For comparison, we present in Fig.~\ref{fig_clean_zero} the pairing
components for the zero-junction case
($\tilde\chi_1=\tilde\chi_2=0$). In this case, the spatial symmetries
are opposite to that of the $\pi $-junction. We have verified that the
$\pi $ junction has the lower free energy.\cite{eschrig03}

We consider next the Josephson current through the junction.  We
assume identical interface parameters for both contacts, that may
depend on the impact angle $\theta_p$ via a parameter $\mu=|\cos
\theta_p |$.  We obtain for small $\vartheta$, $t_{\uparrow\uparrow}$,
and $t_{\downarrow\uparrow}$ the following expression for the
Josephson current density,
\begin{eqnarray}
J&=&-J_c \sin (\tilde \chi_2-\tilde\chi_1  )
\\
J_c&=&eN_f v_f|\Delta|^2 \pi T \sum_{\epsilon_n>0} 
\left\langle
\frac{\mu \; \vartheta^2 |t_{\downarrow\uparrow}|^2 |t_{\uparrow\uparrow}|^2}{
2\sinh \left(\frac{2\epsilon_n L}{v_f\mu}\right) }
\right\rangle_{\! \! \! \mu}
\frac{\epsilon_n^2}{(\epsilon_n^2+|\Delta|^2)^2}, 
\label{jc}
\end{eqnarray}
where $\langle \cdots \rangle_\mu = \int_0^1 {\rm d}\mu (\ldots )$,
and $N_f$ denotes the density of states of the spin-up electrons at the Fermi level in the half metal.
We point out several differences to the case of the Josephson effect
through a normal metal: (i) The minus sign suggests that the
$\pi$-junction is stable for all temperatures and parameters within
the above approximation. (ii) The Josephson current is proportional to
the square of the spin-mixing angle $\vartheta$ and the squares of
both the tunneling amplitudes $t_{\downarrow\uparrow}$ and
$t_{\uparrow\uparrow}$.  Thus, the {\it magnitude} of the critical
current is much more sensitive to the interface characteristics than
in a usual Josephson junction;  in particular, strong sample-to-sample
fluctuations are expected as the magnitude 
is proportional to $t_{\downarrow\uparrow}$. (iii) The
additional phases $\vartheta_{\downarrow\uparrow}+
\vartheta_{\uparrow\uparrow}$ in $\tilde\chi_{1,2}$ can lead to a
shift of the equilibrium phase difference between the superconductors
except when they are identical for both interfaces of the Josephson
junction. 
(iv) The Josephson current density is a result of both the even-frequency
$p$-wave triplet and the odd-frequency $s$-wave triplet amplitudes,
and neither of them can be neglected.

\begin{figure}[t]
\centerline{
\includegraphics[width=\textwidth]{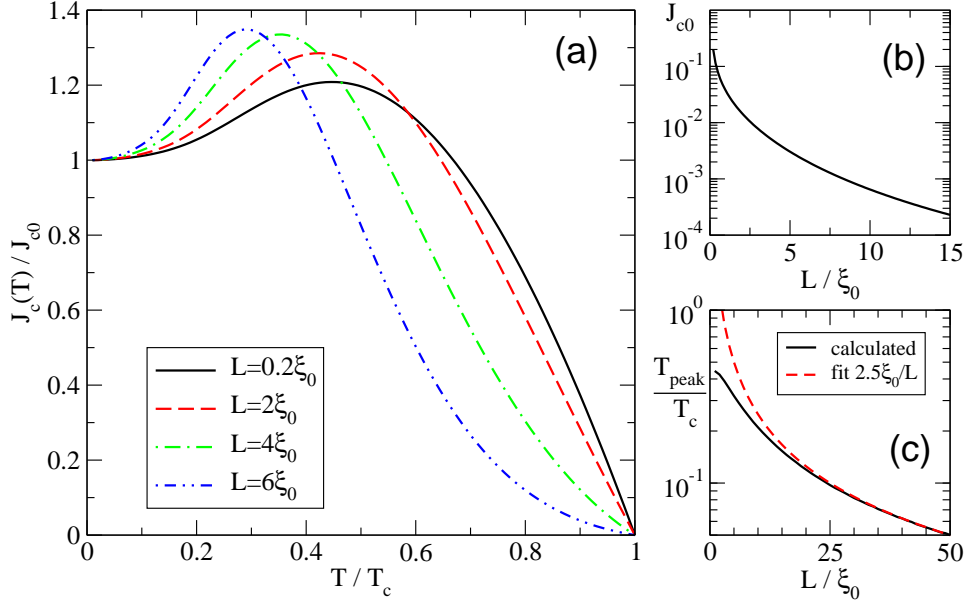}
}
\caption{\small
(a) The normalized critical Josephson current density 
$J_c(T)/J_{c0}$ 
as computed from Eq.~(\ref{jc}) as a function of temperature $T$.
The temperature dependence of the gap is given by the well-known BCS formula.
(b) The zero temperature value $J_{c0}=J_c(T=0)$ is shown in
units of 
$eN_f v_f T_c |t_{\downarrow\uparrow }|^2|t_{\uparrow\uparrow }|^2 \vartheta^2$
as function of junction length $L$.
(c) The peak position of the curves in (a) as function of $L$. 
\label{fig_Jc}
}
\end{figure}

In Fig.~\ref{fig_Jc} (a) we show $J_c(T)$ from Eq.~(\ref{jc})
normalized to its zero temperature value. As above we assumed a variation of
$\theta $, $|\tau_{\uparrow\uparrow}|$, and $|\tau_{\downarrow\uparrow}|$
with impact angle that is proportional to $|\cos \theta_p|$.
As can be seen, there is a
low-temperature anomaly in $J_c(T)$ that has been discussed
previously.\cite{eschrig03} Here we show that this anomaly is a robust
feature that exists even in the limit of small transmission and small
spin-mixing angle, and is independent of these material parameters.
Thus, the dependence on the interface parameters can be divided out
and $J_c(T)/J_c(0)$ is a universal function, only dependent on $L/\xi_0$.
The appearance of the anomaly can be traced back to the different
energy dependence of the pair amplitudes in the half metal compared
with a normal metal, which results in the $\epsilon_n^2$-factor in the
numerator of Eq.~(\ref{jc}).  The position of the peak maximum in the
temperature dependence of the Josephson current is shown in
Fig.~\ref{fig_Jc} (c) as a function of junction length $L$. It scales
at large $L$ as $T_{peak}/T_c\approx 2.5 \xi_0/L$. For small $L$ it
saturates at a finite temperature.  We also show in Fig.~\ref{fig_Jc}
(b) the variation of the zero temperature Josephson current with
junction length. It decays for large $L$, as is $L^{-1}\exp(-L/\xi_0)$.


\section{SUMMARY}

In conclusion, we have investigated the superconducting pairing
correlations with unconventional symmetries at interfaces between
superconductors and ferromagnets. We have demonstrated that
in ballistic superconductor/ferromagnet junctions
spin-active interface scattering naturally leads 
to all possible symmetry components 
of pairing amplitudes compatible with the Pauli principle.
We have also discussed the case of a junction with
a half-metallic ferromagnet. In this case odd-frequency $s$-wave and 
even frequency $p$-wave components are of comparable magnitude and
are essential for the Josephson current. 
This leads to a $\pi$-junction, where the supercurrent is carried by
spin-triplet Cooper pairs.
We have shown that a low-temperature
anomaly in the Josephson effect through a half-metal is of universal
nature in the tunneling limit in ballistic systems.
We have derived analytic expressions for all pairing components and for
the Josephson current in the limit of small interface transmissions and small spin-mixing
angle.


\section*{ACKNOWLEDGEMENTS}
We acknowledge stimulating discussions with H. von L\"ohneysen, 
D. Beckmann, 
V. Chandrasekhar,
G. Goll, 
F. P{\'e}rez-Willard, 
C. S\"urgers, and
H.B. Weber, 
on the properties of superconductor-ferromagnet heterostructures, and
with T. Klapwijk on the observation of triplet supercurrents in a
half-metal.  Our work was also supported by the German-Israeli
Foundation for Scientific Research and Development (J.K.), and the
Alexander von Humboldt Foundation (T.L.).


\end{document}